\documentclass[showpacs,twocolumn,floatfix,superscriptaddress]{revtex4}
\usepackage{graphicx}
\begin{document}
\title{Ehrenfest-time-dependent excitation gap in a chaotic Andreev billiard}
\author{\.{I}. Adagideli and C. W. J. Beenakker}
\affiliation{Instituut-Lorentz, Universiteit Leiden, P.O. Box 9506, 2300 RA Leiden,
The Netherlands}
\date{February 2002}
\begin{abstract}
A semiclassical theory is developed for the appearance of an excitation gap in
a ballistic chaotic cavity connected by a point contact to a superconductor.
Diffraction at the point contact is a singular perturbation in the limit
$\hbar\rightarrow 0$, which opens up a gap $E_{\rm gap}$ in the excitation
spectrum. The time scale $\hbar/E_{\rm gap}\propto\alpha^{-1}\ln\hbar$ (with
$\alpha$ the Lyapunov exponent) is the Ehrenfest time, the characteristic time
scale of quantum chaos.
\end{abstract}
\maketitle

The density of states in a normal metal is suppressed near the Fermi energy
when it is brought into contact with a superconductor. The history of this
proximity effect goes back to the 1960's \cite{Gen63}. It was understood early
on \cite{Mcm68} that the energy range of the suppression is the inverse of the
typical life time $\tau_{\rm c}$ of an electron or hole quasiparticle in the
normal metal. This life time is finite (even at zero temperature) because an
electron is converted into a hole by Andreev reflection at the interface with
the superconductor \cite{And64}. The energy scale $E_{\rm c}=\hbar/\tau_{\rm
c}$, known as the Thouless energy, is the product of the mean level spacing
$\delta$ in the normal metal and the dimensionless conductance of the contact
to the superconductor. (For example, $E_{\rm c}=N\delta$ for coupling via an
$N$-channel ballistic point contact.) The appearance of an excitation gap of
the order of the Thouless energy is the essence of the traditional proximity
effect.

Some years ago it was realized \cite{Mel96,Lod98,Sch99,Pil00,Ihr01,Tar01} that
the proximity effect is essentially different if the normal metal becomes so
small and clean that scattering by impurities can be neglected. This applies to
a quantum dot in a two-dimensional electron gas \cite{Wee97}, and because of
the resemblance to a  billiard (cf.\ Fig.\ 1) one speaks of an ``Andreev
billiard'' \cite{Kos95,Wie02}. Depending on the shape of the billiard, the
classical dynamics varies between integrable and chaotic. No excitation gap is
induced by the proximity effect in an integrable billiard \cite{Mel96,Ihr01}.
An excitation gap does appear in a chaotic billiard \cite{Mel96,Sch99}, but its
magnitude is only given by the Thouless energy if the chaos sets in
sufficiently rapidly \cite{Lod98,Tar01}.

The characteristic time scale of quantum chaos is the Ehrenfest time $\tau_{\rm
E}=\alpha^{-1}\ln(L/\lambda_{\rm F})$, defined in terms of the Lyapunov
exponent $\alpha$ (being the rate at which nearby trajectories diverge
exponentially in time) and the relative magnitude of the Fermi wave length
$\lambda_{\rm F}=2\pi/k_{\rm F}$ and a typical dimension $L$ of the billiard
\cite{Ale96}. Chaotic dynamics requires $\alpha^{-1}\ll\tau_{\rm c}$, but
$\tau_{\rm E}$ could be either smaller or larger than $\tau_{\rm c}$. In the
regime $\tau_{\rm E}\ll\tau_{\rm c}$ the excitation gap is set as usual by the
Thouless energy. Established techniques (random-matrix theory, non-linear
$\sigma$-model) provide a complete description of this regime
\cite{Mel96,Vav01,Ost01,Lam01}. The opposite regime $\tau_{\rm E}\gg\tau_{\rm
c}$ has no analog in the conventional proximity effect. Random-matrix theory is
helpless and this regime has also shown a frustrating resilience to solution by
means of the ballistic $\sigma$-model \cite{Tar01}. In particular, no mechanism
has yet been demonstrated to produce the hard gap at $\hbar/\tau_{\rm E}$
conjectured by Lodder and Nazarov \cite{Lod98}.

\begin{figure}
\includegraphics[width=8cm]{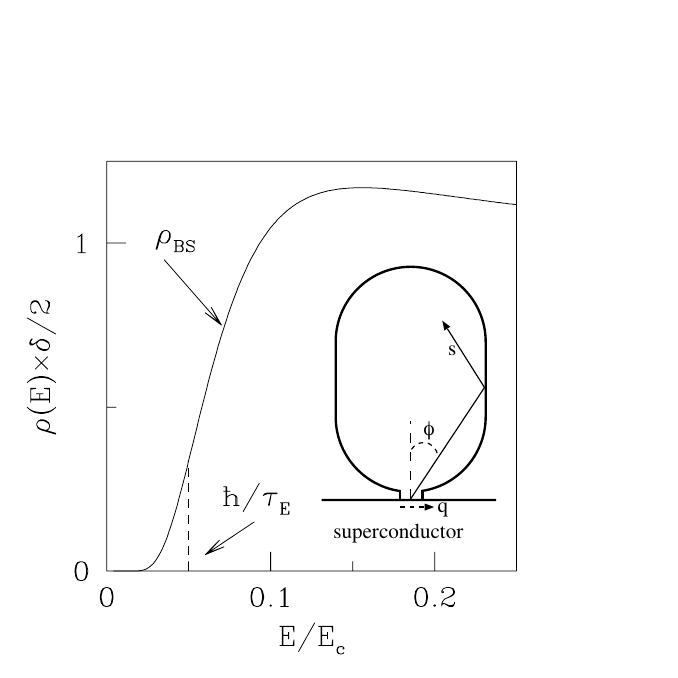}
\caption[]{
Solid curve: Density of states $\rho_{\rm BS}$ of a chaotic Andreev billiard
(inset), which is gapless according to the semiclassical Bohr-Sommerfeld
approximation (\protect\ref{rhoBS}). The dashed line indicates schematically
the phenomenon that we seek to describe in this paper: The opening of a gap at
the inverse Ehrenfest time as a result of diffraction at the contact with the
superconductor.
\label{figrhoBSdot}
}
\end{figure}

Here we report an attack on this problem by an alternative approach, starting
from the semiclassical Andreev approximation to the Bogoliubov-De Gennes (BdG)
equation \cite{And64}. The limit $\tau_{\rm E}\rightarrow\infty$ yields the
Bohr-Sommerfeld approximation to the density of states
\cite{Mel96,Lod98,Sch99},
\begin{equation}
\rho_{\rm BS}(E)=\frac{2}{\delta}\,\frac{(E_{\rm c}/4E)^{2}\cosh (E_{\rm
c}/4E)}{\sinh^{2} (E_{\rm c}/4E)},\label{rhoBS}
\end{equation}
which is gapless (cf.\ Fig.\ \ref{figrhoBSdot}). We have found that diffraction
at the contact with the superconductor is a singular perturbation to $\rho_{\rm
BS}$ that opens up a gap at the inverse Ehrenfest time, and provides an
intuitively appealing mechanism for the gap phenomenon.

We recall the basic equations. The electron and hole components $u({\bf r})$
and $v({\bf r})$ of the spinor wave function satisfy the BdG equation
\begin{equation}
\left(\begin{array}{cc}
H&\Delta\\
\Delta^{\ast}&-H
\end{array}\right)
\left(\begin{array}{c}
u\\
v
\end{array}\right)
=E
\left(\begin{array}{c}
u\\
v
\end{array}\right),
\label{BdG}
\end{equation}
which contains the single-particle Hamiltonian $H=-\nabla^{2}+V({\bf r})-E_{\rm
F}$ (with confining potential $V$) and the pair potential $\Delta({\bf r})$
(vanishing in the normal metal and equal to $\Delta_{0}$ in the
superconductor). The energy $E$ is measured relative to the Fermi energy
$E_{\rm F}=k_{\rm F}^{2}$, in units such that $\hbar^{2}/2m\equiv 1$. (In these
units the mean level spacing $\delta$ is related to the area ${\cal A}$ of the
billiard by $\delta=4\pi/{\cal A}$.) We assume that the motion inside the
billiard is ballistic ($V=0$) and that the interface with the superconductor is
a ballistic point contact of width $W\gg\lambda_{\rm F}$ (so that the number of
channels $N=2W/\lambda_{\rm F}\gg 1$ and the Thouless energy $E_{\rm
c}=N\delta\gg\delta$). We work in the regime $\Delta_{0}\gg\hbar v_{\rm F}/W$
(which also implies $\Delta_{0}\gg E_{\rm c}$), to ensure that the excitation
spectrum is independent of the properties of the superconductor.

For a semiclassical description one substitutes
$(u,v)=(\bar{u},\bar{v})Ae^{iS}$, with $\hbar S$ the action along a classical
trajectory at the Fermi energy. The wave amplitude $A$ is related to the
classical action by the continuity equation $\nabla\cdot(A^{2}\nabla S)=0$,
while $S$ itself satisfies the Hamilton-Jacobi equation $|\nabla S|^{2}=E_{\rm
F}-V$ (so that $\hbar\nabla S$ is the momentum along the trajectory). The BdG
equation takes the form
\begin{equation}
\left(\begin{array}{cc}
-2ik_{\rm F}\partial_{s}+\delta H&\Delta\\
\Delta^{\ast}&2ik_{\rm F}\partial_{s}-\delta H
\end{array}\right)
\left(\begin{array}{c}
\bar{u}\\
\bar{v}
\end{array}\right)
=E
\left(\begin{array}{c}
\bar{u}\\
\bar{v}
\end{array}\right),
\label{BdG1}
\end{equation}
with $\delta H\bar{u}=-A^{-1}\nabla^{2}(A\bar{u})$. The derivative
$\partial_{s}=k_{\rm F}^{-1}(\nabla S)\cdot\nabla$ is taken along the classical
trajectory. The Andreev approximation consists in neglecting the term $\delta
H$ containing second derivatives of the slowly varying functions
$A,\bar{u},\bar{v}$.

We consider a classical trajectory that starts as an electron at a point
$q\in(0,W)$ along the interface with the superconductor, making an angle
$\phi\in(-\pi/2,\pi/2)$ with the normal (cf.\ Fig.\ \ref{figrhoBSdot}). The
product $b=q\cos\phi$ is the ``impact parameter''. The trajectory returns to
the interface after a path length $\ell$, and then it is retraced in the
opposite direction as a hole . The coordinate $s\in(0,\ell)$ runs along one
repetition of this trajectory. We count trajectories with measure
$dq\,d\sin\phi=db\,d\phi$, corresponding to a uniform measure in phase space.
Equivalently, we can sum over scattering channels $n=1,2,\ldots N$, related to
$\phi$ by $n\approx N|\sin\phi|$.

If we ignore the term $\delta H$ in Eq.\ (\ref{BdG1}) we recover the
Bohr-Sommerfeld density of states \cite{Mel96,Lod98,Sch99}. Indeed, without
$\delta H$ the solution of the eigenvalue problem is
\begin{equation}
\left(\begin{array}{c}
\bar{u}_{m}\\
\bar{v}_{m}
\end{array}\right)
=\frac{1}{\sqrt{2\ell}}
\left(\begin{array}{c}
e^{im\pi s/2\ell}\\
ie^{-im\pi s/2\ell}
\end{array}\right),\;\;
E_{m}=m\pi k_{\rm F}/\ell,\label{zerothsolution}
\end{equation}
with $m=\pm 1,\pm 3,\pm 5\ldots$ running over positive and negative odd
integers. The path length $\ell$ in a chaotic billiard varies in a quasi-random
way upon varying the initial conditions $q$ and $\phi$, with an exponential
distribution $P(\ell)=\bar{\ell}^{-1}\exp(-\ell/\bar{\ell})$. (The mean path
length is $\bar{\ell}=4\pi k_{\rm F}/E_{\rm c}$.) The density of states
\begin{equation}
\rho(E)=N\int_{0}^{\infty}d\ell\,P(\ell)
\sum_{m=1,3,5}^{\infty}\delta[E-E_{n}(\ell)]\label{rhoBS1}
\end{equation}
then evaluates to the $\rho_{\rm BS}$ of Eq.\ (\ref{rhoBS}).

The key assumption that will enable us to go beyond the Andreev and
Bohr-Sommerfeld approximations is to assume that the amplitude $A$ varies more
slowly in space than the spinor components $\bar{u}$ and $\bar{v}$, so that we
can approximate $\delta H$ by $-\nabla^{2}$ (neglecting derivatives of $A$).
The characteristic length scale $L_{A}$ for the spatial dependence of $A$ is
set by the smoothness of the confining potential $V$, while the characteristic
length scale for $\bar{u},\bar{v}$ is the contact width $W$. By assuming
$L_{A}\gg W$ we consider the case that diffraction occurs predominantly at the
interface with the superconductor, rather than inside the billiard. Since $A$
depends on the shape of the billiard, this is the regime in which we can hope
to obtain a geometry-independent ``universal'' result.

To investigate the effect of $\delta H$ we restrict the dimensionality of the
Hilbert space in two ways: Firstly, we neglect any mixing of the $N$ scattering
channels. (This is known to be a good approximation of the diffraction that
occurs when a narrow constriction opens abruptly into a wide region
\cite{Sza89}; it does not require smooth corners in the contact.) Secondly,
since we are interested in excitation energies $E\ll E_{\rm c}$, we include
only the two lowest eigenstates $m=\pm 1$ of the zeroth-order solution
(\ref{zerothsolution}). [The contributions from higher levels are smaller by a
factor $\exp(-E_{\rm c}/2E)$.] We need to include both $E_{1}$ and $E_{-1}$,
although the excitation spectrum contains only positive eigenvalues, because of
the (virtual) transitions between these two levels induced by $\delta H$. With
these restrictions we have for each scattering channel a one-dimensional
eigenvalue problem. The effective Hamiltonian ${\cal H}_{\rm eff}$ is a
$2\times 2$ matrix differential operator acting on functions of $b$.

We write ${\cal H}_{\rm eff}={\cal H}_{0}+{\cal H}_{1}$, where ${\cal H}_{0}$
corresponds to the Andreev approximation and ${\cal H}_{1}$ contains the
diffractive effects. The zeroth-order term is diagonal,
\begin{equation}
{\cal H}_{0}=
\left(\begin{array}{cc}
\pi k_{\rm F}/\ell(b)&0\\
0&-\pi k_{\rm F}/\ell(b)
\end{array}\right).
\label{H0}
\end{equation}
The relation between $\ell$ and $b$ is determined by the differential equation
$d\ell/db=g(b)\exp(\kappa\ell)$, which expresses the exponential divergence of
nearby trajectories (in terms of a Lyapunov exponent $\kappa=\alpha/v_{\rm F}$
given as inverse length rather than inverse time). The pre-exponential $g(b)$
is of order unity, changing sign at extrema of $\ell(b)$. Upon integration one
obtains
\begin{equation}
\kappa\ell(b)=-\ln|\kappa b|+{\cal O}(1),\label{lbrelation}
\end{equation}
where we have shifted the origin of $b$ such that $b=0$ corresponds to a local
maximum $\ell_{\rm max}\gg\bar{\ell}$ of $\ell(b)$. [The logarithmic
singularity is cut off at $|\kappa b|\lesssim\exp(-\kappa\ell_{\rm max})$.]
There is an exponentially large number ${\cal
N}(\ell)\propto\exp(\kappa\ell-\ell/\bar{\ell})$ of peaks around which Eq.\
(\ref{lbrelation}) applies.

To obtain the diffractive correction ${\cal H}_{1}$, in the regime that $\delta
H=-\nabla^{2}$, we express the Lapacian in the local displacements $ds$ and
$d\zeta=e^{\kappa s}db$. Since these are approximately orthogonal for $\kappa
s\lesssim 1$, we have
\begin{equation}
\delta H=-\partial_{s}^{2}-e^{-2\kappa s}\partial_{b}^{2}.\label{nabladef}
\end{equation}
The first term $\partial_{s}^{2}$ is a small correction to the zeroth order
density of states. The second term $\partial_{b}^{2}$, in contrast, is a
singular perturbation because it associates a kinetic energy with the variable
$b$. The resulting zero-point motion implies a non-zero ground state energy,
and hence it is responsible for the opening of an excitation gap. Projecting
${\cal H}_{1}$ onto the space spanned by the two lowest eigenfunctions $n=\pm
1$ of Eq.\ (\ref{zerothsolution}), and retaining only the leading order terms
in $1/\kappa\ell$, we find
\begin{equation}
{\cal H}_{1}=
\left(\begin{array}{cc}
0&i\\
-i&0
\end{array}\right)
\frac{\pi}{4}
\frac{d}{db}[\kappa\ell(b)]^{-2}\frac{d}{db}+{\cal O}(\kappa\ell)^{-3}.
\label{H1}
\end{equation}

\begin{figure}
\includegraphics[width=8cm]{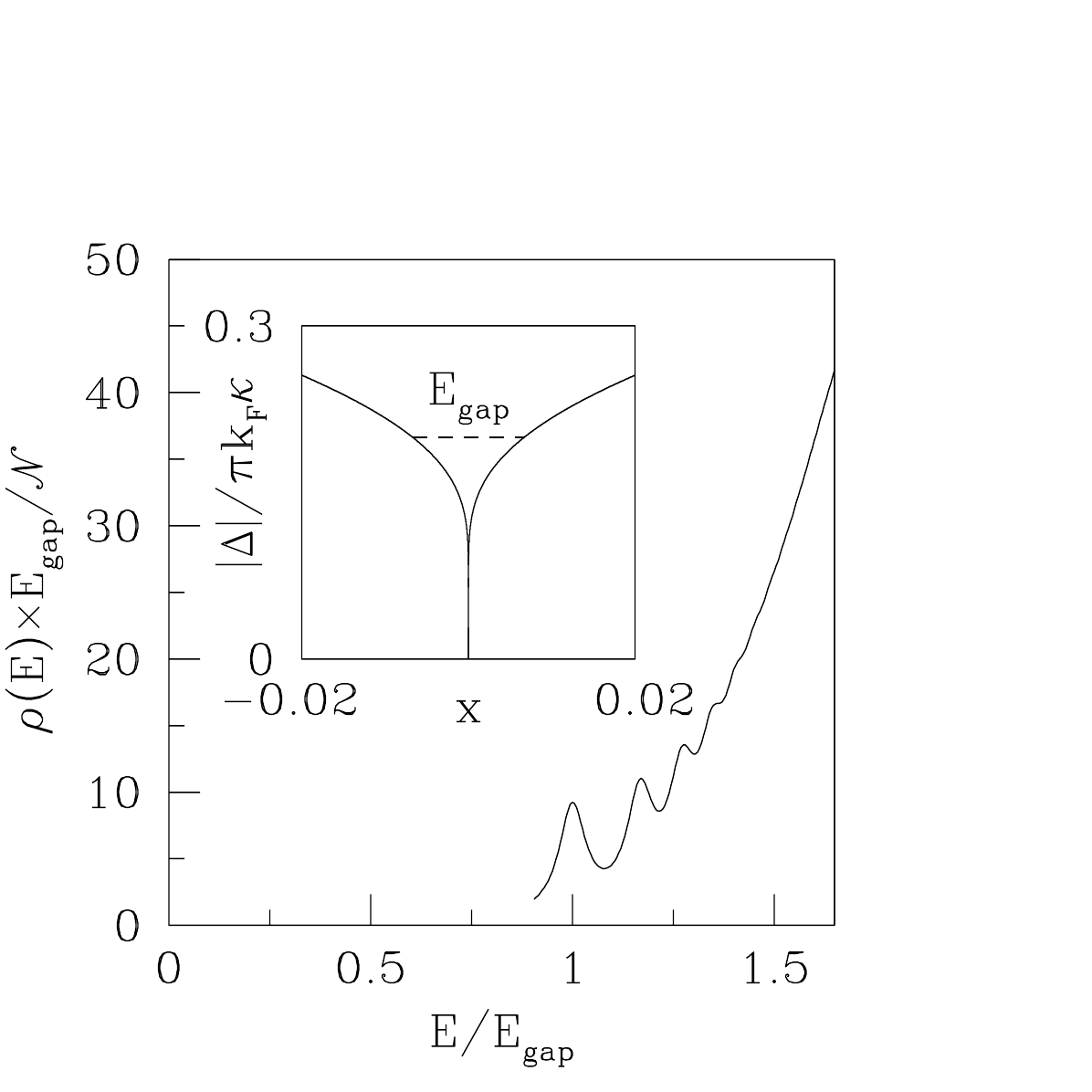}
\caption[]{
Low-energy density of states $\rho(E)$ of the effective Hamiltonian
(\protect\ref{Heff}), related to $\rho(\omega)$ of the biharmonic equation
(\protect\ref{biharmonic3a}) by Eq.\ (\protect\ref{rhoEomega}). The plot is for
$|\ln\epsilon|=10$ and has been smoothed with a Lorentzian. The inset shows the
logarithmic pair potential appearing in ${\cal H}_{\rm eff}$, the ground state
of which is the excitation gap (dashed line).
\label{figrhoE}
}
\end{figure}

The effective Hamiltonian can be brought into a more familiar form by the
unitary transformation ${\cal H}_{\rm eff}\rightarrow
e^{-i\sigma_{1}\pi/4}{\cal H}_{\rm eff}e^{i\sigma_{1}\pi/4}$ (with $\sigma_{i}$
a Pauli matrix), followed by the change of variable $x=\kappa b-\kappa
b\ln|\kappa b|$ (in the range $|x|<1$).  We work again to leading order in
$1/\kappa\ell$, and find
\begin{equation}
{\cal H}_{\rm eff}=
\pi k_{\rm F}\kappa\left(\begin{array}{cc}
-\epsilon\,\partial_{x}^{2}&-i/\ln|x|\\
i/\ln|x|&\epsilon\,\partial_{x}^{2}
\end{array}\right),\;\;\epsilon\equiv\frac{\kappa}{4k_{\rm F}}.
\label{Heff}
\end{equation}
This effective Hamiltonian has the same form as the BdG Hamiltonian
(\ref{BdG}), for a fictitious one-dimensional system having $V=E_{\rm F}$ and
having a pair potential $\Delta(x)$ that vanishes logarithmically $\propto
1/\ln|x|$ at the origin (cf.\ Fig.\ \ref{figrhoE}). The kinetic energy
$\epsilon\,\partial_{x}^{2}$ gives a finite excitation gap, even though
$\epsilon\ll 1$. Let us now compute this gap.

Since $e^{i\sigma_{2}\pi/4}{\cal H}_{\rm eff}^{2}e^{-i\sigma_{2}\pi/4}$ is a
diagonal matrix, the spectrum of ${\cal H}_{\rm eff}$ is given by the scalar
eigenvalue problem
\begin{equation}
\left|\epsilon\frac{d^{2}}{dx^{2}}+\frac{i}{\ln|x|}\right|^{2}\Psi(x)=
\left(\frac{E}{\pi k_{\rm F}\kappa}\right)^{2}\Psi(x).\label{biharmonic}
\end{equation}
The ground state energy is the excitation gap $E_{\rm gap}$. To generate an
asymptotic expansion of $E_{\rm gap}$ for small $\epsilon$, we first multiply
both sides of Eq.\ (\ref{biharmonic}) by a factor $Z^{2}$ and then substitute
$x=X\sqrt{\epsilon Z}$. This results in
\begin{eqnarray}
&&\left|\frac{d^{2}}{dX^{2}}+iU\right|^{2}\Psi=\left(\frac{ZE}{\pi k_{\rm
F}\kappa}\right)^{2}\Psi,\label{biharmonic2a}\\
&&U(X)=\frac{2Z}{\ln\epsilon Z}\left[1-\frac{2\ln|X|}{\ln\epsilon Z}+{\cal
O}(\ln\epsilon Z)^{-2}\right].\label{biharmonic2b}
\end{eqnarray}
We now choose $Z$ such that $Z^{2}=-\ln^{3}\epsilon Z$ and obtain the
biharmonic equation
\begin{eqnarray}
&&(d^{4}/dX^{4}+16\ln|X|)\Psi=\omega\Psi,\label{biharmonic3a}\\
&&\omega=(ZE/\pi k_{\rm F}\kappa)^{2}-4Z^{2/3}+{\cal
O}(Z^{-1/3}).\label{biharmonic3b}
\end{eqnarray}

The ground state of Eq.\ (\ref{biharmonic3a}) is at $\omega_{0}=14.5$.
Substituting in Eq.\ (\ref{biharmonic3b}), and using
$Z^{2/3}=|\ln\epsilon|-\frac{3}{2}\ln|\ln\epsilon|+{\cal O}(1/\ln\epsilon)$, we
arrive at
\begin{equation}
E_{\rm gap}=\frac{2\pi k_{\rm
F}\kappa}{|\ln\epsilon|}\left(1+\frac{3\ln|\ln\epsilon|}{2|\ln\epsilon|}+
\frac{\omega_{0}}{8|\ln\epsilon|}+{\cal
O}(\ln\epsilon)^{-3/2}\right).\label{Egapresult}
\end{equation}
Only the leading order term is significant in view of the approximations made
in Eq.\ (\ref{Heff}). Restoring the original variables we have
\begin{equation}
E_{\rm gap}=\frac{\pi\hbar\alpha}{\ln(v_{\rm F}/\alpha\lambda_{\rm
F})}.\label{Egapresult2}
\end{equation}
The Ehrenfest time $\tau_{\rm E}=\alpha^{-1}\ln(L/\lambda_{\rm F})$ contains
the classical length $L=v_{\rm F}/\alpha$, which is of the order of the linear
dimension of the billiard.

The density of states $\rho(\omega)$ of the biharmonic equation
(\ref{biharmonic3a}) can be calculated numerically \cite{Gre97}. The density of
states $\rho(E)$ near the gap is related to $\rho(\omega)$ by
\begin{equation}
\rho(E)=\frac{8{\cal N}|\ln\epsilon|}{E_{\rm
gap}}\rho[\omega=\omega_{0}+8|\ln\epsilon|(E/E_{\rm gap}-1)],\label{rhoEomega}
\end{equation}
and is plotted in Fig.\ \ref{figrhoE} for $|\ln\epsilon|=10$. The factor ${\cal
N}\propto\exp(\pi k_{\rm F}\kappa/E_{\rm gap}-E_{\rm c}/E_{\rm gap})$ counts
the number of peaks in $\ell(b)$ around which ${\cal H}_{\rm eff}$ applies. The
Bohr-Sommerfeld approximation (\ref{rhoBS}) corresponds to the large-$\omega$
asymptote $\rho(\omega)=\frac{1}{16}\exp(\omega/16)$. Since
$\omega-\omega_{0}\gg 1$ implies $E/E_{\rm gap}-1\gg 1/|\ln\epsilon|$, the
width $\Delta E\simeq E_{\rm gap}/|\ln\epsilon|$ of the energy range above the
gap in which the Bohr-Sommerfeld approximation breaks down is small compared to
the gap itself.

Because ${\cal H}_{\rm eff}$ has only a few levels in the range $\Delta E$, the
density of states $\rho(E)$ oscillates strongly in this range. These levels are
highly degenerate (by a factor ${\cal N}$) in our approximation. Tunneling
between the levels will remove the degeneracy and smooth the oscillations. (A
small amount of smoothing has been inserted by hand in Fig.\ \ref{figrhoE}.) We
surmise that some oscillatory energy dependence will remain, but this is an
aspect of the problem that needs further study.

In conclusion, we have analyzed a mechanism for the ``gap phenomenon'' in the
proximity effect of chaotic systems. Diffraction at the contact with the
superconductor is described by an effective Hamiltonian ${\cal H}_{\rm eff}$
that contains (1) a kinetic energy which vanishes in the classical limit and
(2) a pair potential with a logarithmic profile. The resulting excitation gap
$E_{\rm gap}$ (being the ground state energy of ${\cal H}_{\rm eff}$) vanishes
logarithmically as the ratio of the Fermi wavelength and a classical length
scale (set by the Lyapunov exponent) goes to zero. The time scale $\hbar/E_{\rm
gap}$ is the Ehrenfest time, providing a manifestation of quantum chaos in the
superconducting proximity effect.

This work was supported by the Dutch Science Foundation NWO/FOM. We thank A. N.
Morozov and P. G. Silvestrov for helpful discussions.

\end{document}